\pdfoutput=1
\documentclass[11pt,twoside,a4paper,cmspaper,final,collab]{cms-tdr}
\def\svnVersion{65063f8}\def\svnDate{2026/11/09}\def\cmsCernNoTag{CERN-EP-2025-250}\def\cmsCernDate{\today}\def\cmsMessage{Published in Physics Letters B as \href{http://dx.doi.org/10.1016/j.physletb.2026.141066}{\doi{10.1016/j.physletb.2026.141066}.}}
\begin{document}\cmsNoteHeader{HIN-24-006}

\newcommand{\minv}{\ensuremath{m_{\text{inv}}}\xspace}
\newcommand{\vn}{\ensuremath{v_{\mathrm{n}}}\xspace}
\newcommand{\vtwo}{\ensuremath{v_{2}}\xspace}
\newcommand{\vthree}{\ensuremath{v_{3}}\xspace}
\newcommand{\Dstophipi}{\ensuremath{\PDpms \to\PGfP{1020}+\PGppm \to(\PKp+\PKm)+\PGppm}}
\newcommand{\PbPb}{\ensuremath{\text{PbPb}}\xspace}
\newcommand{\QnA}{\ensuremath{Q_{n \mathrm{A}}}\xspace}
\newcommand{\QnB}{\ensuremath{Q_{n \mathrm{B}}}\xspace}
\newcommand{\QnC}{\ensuremath{Q_{n \mathrm{C}}}\xspace}
\newcommand{\wj}{\ensuremath{w_{j}}\xspace}
\ifthenelse{\boolean{cms@external}}{\providecommand{\cmsTable}[1]{\resizebox{\columnwidth}{!}{#1}}}{\providecommand{\cmsTable}[1]{#1}}

\newlength\cmsTabSkip\setlength{\cmsTabSkip}{1ex}

\cmsNoteHeader{HIN-24-006}
\title{Influence of strangeness on the anisotropic flow of prompt \PDpms mesons in PbPb collisions at \texorpdfstring{$\sqrtsNN = 5.02\TeV$}{sqrtsNN = 5.02 TeV}}

\date{\today}

\abstract{
The azimuthal anisotropy of prompt \PDpms mesons produced in lead-lead (PbPb) collisions at a nucleon-nucleon center-of-mass energy of 5.02\TeV is measured using data obtained with the CMS detector. The dataset corresponds to an integrated luminosity of 0.58\nbinv. The azimuthal anisotropy of heavy charmed mesons provides a key constraint on the interactions of charm quarks with the quark-gluon plasma (QGP) medium. These interactions include coalescence mechanisms and parton energy loss in the QGP. The anisotropy is characterized by measurements of the second- (\vtwo) and third-order (\vthree) Fourier coefficients of the azimuthal distribution of the \PDpms mesons. The \vtwo coefficient is determined in the transverse momentum range $4 < \pt < 40\GeV$ for three event centrality classes, while the \vthree coefficient is measured in the range $4 < \pt < 20\GeV$ for a single event centrality class. The results for the \PDpms mesons are compared to those previously measured for \PDz mesons. For their respective \pt and centrality ranges, the \vtwo and \vthree coefficients for \PDpms and \PDz mesons are found to be consistent. This suggests that the strangeness content of the \PDpms meson does not significantly alter its azimuthal distribution, at least within the measured \pt ranges. Theoretical predictions incorporating heavy-quark transport and hadronization mechanisms in the QGP are compared to the measured anisotropy coefficients. 
}

\hypersetup{%
pdfauthor={CMS Collaboration},%
pdftitle={Influence of strangeness on the anisotropic flow of prompt Ds mesons in PbPb collisions at sqrt(sNN) = 5.02 TeV},%
pdfsubject={CMS},%
pdfkeywords={CMS, heavy ion, heavy flavor, flow}}

\maketitle

\section{Introduction}
\label{sec:intro}

In ultrarelativistic heavy ion collisions, cold nuclear matter undergoes a transformation into a dense, strongly interacting state, known as quark-gluon plasma (QGP). 
This transformation releases quarks and gluons from their hadronic confinement~\cite{BRAHAMS:2005,PHOBOS:2005,STAR:2005,PHENIX:2005,cms2025overview,QGP_ref_1,QGP_ref_2}. 
A notable feature of QGP is the collective motion of its constituents. 
Heavy ion collisions result in anisotropic pressure gradients in the QGP that cause an azimuthal anisotropy in the emitted particles. The anisotropy is quantified using Fourier coefficients ($v_n$) of the azimuthal distribution to characterize this collective flow. This phenomenon represents a non-trivial collective response where the initial-state spatial eccentricities ($\varepsilon_n$) of the overlap region are transferred via pressure gradients in the thermalized medium into a final-state anisotropy in momentum space. 
The observed $v_n$ values for light hadrons having transverse momenta $0.3<\pt<3\GeV$ are consistent with predictions from relativistic 
hydrodynamics~\cite{Ollitrault:1992bk,Heinz:2013th,Gale:2013da}. The second coefficient, \vtwo, referred to as ``elliptic flow," results from the interplay between event-by-event fluctuations in the initial-state density distribution of the QGP medium and the overall lenticular shape of this distribution for noncentral collisions. The third coefficient, \vthree, or ``triangular flow," mainly arises from fluctuations in the initial-state geometry~\cite{Alver:2010prc}.

As a result of their substantial mass, charm and bottom quarks are produced early in the collision process~\cite{Prino:2016hf}, with their azimuthal distributions possibly influenced by interactions with the QGP they traverse~\cite{Beraudo:2018enl}. In the low transverse momentum regime, partons are expected to undergo frequent interactions within the medium, leading to thermalization and a collective flow pattern. Heavy quarks also participate in this collective expansion, with charm quarks, in particular, expected to follow the flow of the light-flavor partons~\cite{vanHes:2006therm}. 
In the high-\pt range (${>} 3\GeV$), an azimuthal asymmetry may result from path length dependent parton energy loss~\cite{Gyulassy2001hptv2,Shuryak2002hptv2}. 
Hence, investigating the flow coefficients of heavy quarks in heavy ion collisions offers valuable insight into key properties of the QGP, such as its transport coefficients, the degree of charm quark thermalization, and the path length dependence of parton energy loss through this medium. As quarks are not directly observable, the flow of heavy quarks is inferred from the measured flow of heavy hadrons. Although the \vtwo and \vthree coefficients for charm hadrons have been extensively studied in lead-lead (\PbPb) collisions at the CERN LHC~\cite{Acharya:2018dvn,Sirunyan:2017plt,Sirunyan:2020dvn,Aad:2020bvn,Acharya:2020jpsivn,Acharya:2021Dvn,Sirunyan:2022dv24, tumasyan2023measurements}, comparing measurements for different charm mesons is important to fully understand how  charm quarks interact with the medium. In particular, comparing the flow behavior of \PDpms mesons, which contain a strange quark, to that of \PDz mesons provides unique sensitivity to the consequence of strangeness enhancement in the QGP. Because of the high energy densities associated with its creation, the QGP medium is expected to have an enhanced strange quark fraction compared to proton-proton collisions~\cite{strangeQGP-1,strangeQGP-2}. In studying heavy quark flow, it is also important to explore how the passage of heavy quarks through the QGP influences the subsequent hadronization processes. For instance, a charm quark might have too high an energy to become thermalized. However, during hadronization via quark coalescence, if this heavy quark recombines with a thermalized light quark from the expanding medium, the resultant charm meson can develop an amplified in-plane flow behavior inherited from the constituent light quark~\cite{rapp2005,he2012}. This effect is strongly mass-dependent, as heavier bottom quarks are not observed to develop significant collective flow via this recombination mechanism~\cite{PhysRevLett.123.192301}. 

In this Letter, the \vtwo and \vthree of prompt \PDpms mesons, that is, mesons produced at the primary vertex containing both a charm and a strange quark, are measured using PbPb collisions. The achieved precision allows for the first detailed comparisons with non-strange charm hadrons. The data were obtained at a center-of-mass energy per nucleon pair of $\sqrtsNN = 5.02 \TeV$ using the CMS detector at the CERN LHC. 
The \vtwo coefficient for \PDpms mesons within the rapidity range $\abs{y} < 2.4 $ is determined as a function of \pt, covering the range 4--40\GeV, in the 0--10\% (central), 10--30\% (mid-central) and 30--50\% (mid-peripheral) centrality classes. The \vthree coefficient is also measured in the 10--30\% centrality class for the \pt range 4--20\GeV. These centrality percentages represent the fraction of the total inelastic hadronic cross section, with 0\% corresponding to full overlap of the colliding nuclei. The broad range of \pt values covered allows for the \PDpms meson azimuthal dependence to be investigated in regions where different physics processes are expected to contribute. Charm thermalization and recombination processes are expected to be important at the lower \pt values studied. In contrast to this, the path-length dependent energy loss of the charm quark traveling through the QGP medium is expected to dominate the $v_n$ coefficients seen at higher \pt values. While the low \pt $v_n$ values serve as a signature of collective flow, the anisotropy at high \pt arises from the geometry-dependent suppression of hard-scattered partons, rather than the collective expansion of the bulk medium. To gain insight into how the \PDpms meson azimuthal asymmetry might be affected by the interactions with the medium, a comparison is made with azimuthal anisotropy measurements of non-strange \PDz mesons~\cite{Sirunyan:2020dvn}. Tabulated results are provided in the HEP Data record for this analysis~\cite{HEPData}.

\section{Experimental setup and data sample}
\label{sec:exp}

The CMS apparatus~\cite{Chatrchyan:2008zzk} is a multipurpose, nearly hermetic detector,
designed to trigger on~\cite{CMS:2020cmk,CMS:2016ngn} and identify electrons,
muons, photons, and hadrons~\cite{CMS:2020uim,CMS:2018rym,CMS:2014pgm}. To determine the azimuthal asymmetry of emitted particles, the information from two subdetectors is used. A multi-layer silicon pixel and strip detector tracker measures charged particle trajectories within the pseudorapidity range $\abs{\eta}<3.0$. Extending the $\eta$ coverage, two hadronic forward (HF) calorimeters, one located on either side of the interaction point, span the range $3.0 < \abs{\eta} < 5.0$. These calorimeters are made of steel and quartz fibers and are segmented to form $0.175{\times}0.175$ ($\Delta\eta{\times}\Delta\phi$) ``towers," where $\phi$ is the azimuthal angle in radians. Event selection is carried out through a two-level trigger system.
The first level employs dedicated hardware processors that analyze signals from the calorimeters and muon detectors~\cite{CMS:2020cmk}. The second level, referred to as the high-level trigger~\cite{CMS:2016ngn}, uses a computing farm executing a streamlined version of the full event reconstruction algorithm to characterize events.

The analyzed Run2 data, collected in 2018, consist of 4.27 billion events selected with a minimum bias (MB) trigger, corresponding to an integrated luminosity of 0.58\nbinv~\cite{CMS:2021xjt, cmsPbPbLumi5p02TeV}. The MB trigger requires energy above selection thresholds in the range of 6--12\GeV 
in both of the HF calorimeters~\cite{CMS:2016ngn}. To suppress background events such as beam-gas interactions and non-hadronic collisions, a number of trigger requirements and offline event selection filters are applied, as described in Ref.~\cite{Sirunyan:2020dvn}.
The final event selection requires the presence of at least one reconstructed vertex based on two or more tracks and positioned within 15\cm of the nominal interaction point along the beam axis. The primary vertex used in the analysis is chosen as the reconstructed vertex with the largest number of associated tracks. Additionally, a coincidence filter is used that requires the presence of at least two calorimeter towers having a detected energy above 4 \GeV in each of the HF detectors. This requirement not only ensures a genuine hadronic collision but also helps suppress contributions from electromagnetic contamination (largely created by ultra-peripheral collisions~\cite{upc}) and beam-related noise. To further suppress beam-gas and other non-hadronic backgrounds, events with pixel cluster shapes inconsistent with the expectations for single PbPb collisions are excluded using a cluster compatibility filter. The centrality is calculated using the transverse energy deposited in the HF calorimeters~\cite{CMS:2011iwn}.

Simulated events from Monte Carlo (MC) generators are used to study the kinematics of \PDpms  mesons in \PbPb collisions. 
The \PDpms mesons are generated with \PYTHIA 8.212~\cite{Sjostrand:2014zea}, tune CP5~\cite{Sirunyan:2019dfx}, 
while their decays are modeled with \EVTGEN 1.3~\cite{Lange:2001uf}. 
Both prompt and non-prompt \PDpms samples are generated. Prompt mesons are produced at the primary vertex, while non-prompt mesons result from \PQb-flavored hadron decays. Prompt mesons are expected to exhibit a larger azimuthal anisotropy compared to non-prompt mesons since the smaller mass of the charm quark compared to the bottom quark allows for a faster thermalisation rate. Also, the decay kinematics of the parent \PQb-hadron reduces the correlation of the daughter \PDpms meson with respect to the bulk medium. The decay products of \PDpms mesons are then embedded into the minimum bias events generated using \HYDJET 1.9~\cite{Lokhtin:2005px} that is tuned to reproduce the charged particle multiplicity and spectra~\cite{Khachatryan:2016odn}. The CMS detector response is simulated with \GEANTfour~\cite{Agostinelli:2002hh}. 

\section{Analysis procedure}
\label{sec:ana}
 
Inclusive (both prompt and non-prompt) \PDpms meson candidates are reconstructed via the decay channel \Dstophipi. 
All tracks used in the analysis have $\pt>1.5\GeV$ and $\abs{\eta}<2.4$. 
Only reconstructed tracks are used that satisfy high-purity selection criteria that impose robust requirements on the track quality~\cite{CMS:2014pgm}.
Each track must be associated with at least 11 hits in the tracker. The track fit quality is constrained by requiring that the $\chi^{2}$ value normalized by both the number of degrees of freedom and the number of tracker layers containing hits along the track, $\chi^2/\text{dof}/N_\text{layers}$, is less than 0.18. Additionally, tracks are required to have a transverse momentum resolution ($\sigma_{\pt}/\pt$) of less than 10\%. The kaon (\PKpm) candidates are reconstructed as charged tracks in the silicon tracker and are selected based on their charge and a kaon mass hypothesis, without the use of dedicated particle identification detectors. 
The \PGfP{1020} candidates are then formed by combining two oppositely charged kaon candidates and are required to have an invariant mass within ${\pm}9\MeV$ of the nominal \PGfP{1020} meson mass of 1019.46\MeV~\cite{Workman:2024pdg}. Finally, the \PDpms candidates are formed by combining \PGfP{1020} candidates with charged tracks to which the \PGppm mass is assigned. The invariant mass ({\minv}) of the reconstructed \PDpms candidates is required to differ by no more than ${\pm}200\MeV$ from the world-average \PDpms meson mass of 1968\MeV~\cite{Workman:2024pdg}. 

Events are further selected by applying a boosted decision tree (BDT) algorithm implemented using the \textsc{tmva} package~\cite{Hocker:2007ht}. 
For the BDT training, the signal candidates are taken from simulated events in which the reconstructed \PDpms mesons are required to match the generated prompt \PDpms particles. The background training is based on a data-derived combinatorial background sample. This sample is composed of \PDpms candidates whose masses are in the range $4-6$ standard deviations ($\sigma$) away from the nominal \PDpms meson mass, with $\sigma$ determined by a fit to the 
observed \PDpms mass peak. Training is performed separately for each \pt range and inclusive centrality class. The variables that are used to discriminate the signal from the background are taken from the mother \PDpms, the \PGfP{1020} mesons, and the final daughter tracks. The variables related to the mother \PDpms meson are:  
the distance between primary and secondary vertices of \PDpms candidates and its uncertainty (decay length significance), the angle between the momentum of the \PDpms candidates and the line joining the primary and secondary vertices of the \PDpms candidates (pointing angle), and the $\chi^{2}$ probability for the \PDpms vertex fit (vertex probability). The variables related to the \PGfP{1020} meson are: the \PGfP{1020} candidate decay length significance, its pointing angle, and the vertex probability. In addition, a variable using the \pt values of the \PGfP{1020} meson decay products is given by $\left|(p_\mathrm{T,1} - p_\mathrm{T,2})/(p_\mathrm{T,1} + p_\mathrm{T,2})\right|$, where $p_\mathrm{T,1}$ and $p_\mathrm{T,2}$ correspond to the \pt values of the $K^\pm$ tracks. These variables are chosen by analyzing their importance in the decision tree and by how they relate to other variables based on the correlation matrix. 
The BDT requirement is optimized to provide the maximal prompt \PDpms meson signal significance (yield over its uncertainty) for each analysis bin. To account for the detector acceptance and efficiency, each \PDpms candidate is further weighted by a \PDpms \pt and centrality dependent factor determined using the simulated event sample. This factor is obtained by taking the ratio of the total number of generated prompt \PDpms mesons in a specified kinematic range to the number of reconstructed \PDpms mesons in that range that satisfy the selection criteria.   

The measurement of \vn coefficients for inclusive \PDpms meson production uses the same scalar-product (SP) method~\cite{STAR:sp, SP_ref} employed in earlier CMS publications on the azimuthal correlations of prompt \PDz mesons~\cite{Sirunyan:2017plt,Sirunyan:2020dvn}. In this approach, separate $Q$-vectors are defined for the \PDpms particles and for unidentified charged particles in three ``reference" subevent ranges. The reference subevents are used to provide $Q$-vectors that represent the event-by-event anisotropy of the collision. For the \PDpms particles, $Q_{n}^{\mathrm{D_{s}^{\pm}}}  \equiv \sum_{j} \re^{in\phi_{j}}$, where $\phi_{j}$ is the azimuthal angle in the laboratory frame of the $j^{th}$ particle and the sum is over all \PDpms candidates with $\abs{\eta}<2.4$ in a given \pt range and centrality class. The corresponding $\vn\{\mathrm{SP}\}$ value is then determined using
\begin{equation}
\label{eq:scalar_product_vn}
v_n \{ \mathrm{SP} \} = \frac{ \langle Q_n^{\mathrm{D_{s}^{\pm}}} Q_\mathrm{nA}^{*} \rangle }{\sqrt{{ \langle Q_\mathrm{nA}^{*}Q_\mathrm{nB} \rangle \langle Q_\mathrm{nA}Q_\mathrm{nC}^{*} \rangle }/{ \langle Q_\mathrm{nB}Q_\mathrm{nC}^{*}\rangle}}}.
\end{equation}
The reference subevent $Q$-vectors are defined as $Q_{n} \equiv \sum_{j=1}^{M} \wj \re^{in\phi_{j}}$, with the sum running over $M$ reference charged particles used to estimate the event-plane orientation in the selected detector region. The asterisk implies complex conjugation, and the subscripts A, B, and C refer to groups of particles in the same event taken from different parts of the detector. Specifically, particles are selected from the three subevent ranges $\mathrm{HF}^{-}$ ($-5 < \eta < -3 $), $\mathrm{HF}^{+}$ ($3 < \eta < 5 $), and the tracker region ($ \abs{\eta} < 0.75 $), respectively. 
In each event, the sum for \QnA and \QnB uses hadrons detected in the respective HF range, while the sum for \QnC includes reconstructed tracks. 
The weight \wj  corresponds to the transverse energy deposited in the HF tower for subevents A and B, and to the track \pt in subevent C, at the angle $\phi_j$. The subevent A (B) corresponds to the HF detector at negative (positive) $\eta$ when the \PDpms meson candidate is at positive (negative) $\eta$.
This choice avoids autocorrelations and results in an $\eta$ gap of at least three units between the \PDpms meson candidates and particles from the underlying subevent, thereby suppressing short-range correlations.
Flattening and recentering procedures are applied to the $Q$-vectors related to the HF and tracker detectors in order to remove acceptance effects~\cite{Poskanzer:1998yz,Alt:2003ab,Qvec_ref}.
The $w_j$ weighted averages $\langle \QnA^{*}\QnB \rangle$, $\langle \QnA\QnC^{*} \rangle$, and $\langle \QnB\QnC^{*} \rangle$ are evaluated using all selected events, while the weighted average $\langle Q_n^{D_{s}^{\pm}} Q_{nA}^{*} \rangle$ is for all selected events where there are \PDpms meson candidates~\cite{Sirunyan:2020dvn,np_D0}. For the nominal $v_n$ results, the 3 dimensional distance of closest approach (DCA) of the \PDpms reconstructed trajectories from the primary vertex is required to satisfy $\mathrm{DCA} <0.01$\cm to enhance the prompt \PDpms fraction. 

\begin{figure*}[htbp!]
    \centering
    \includegraphics[width=0.99\textwidth]{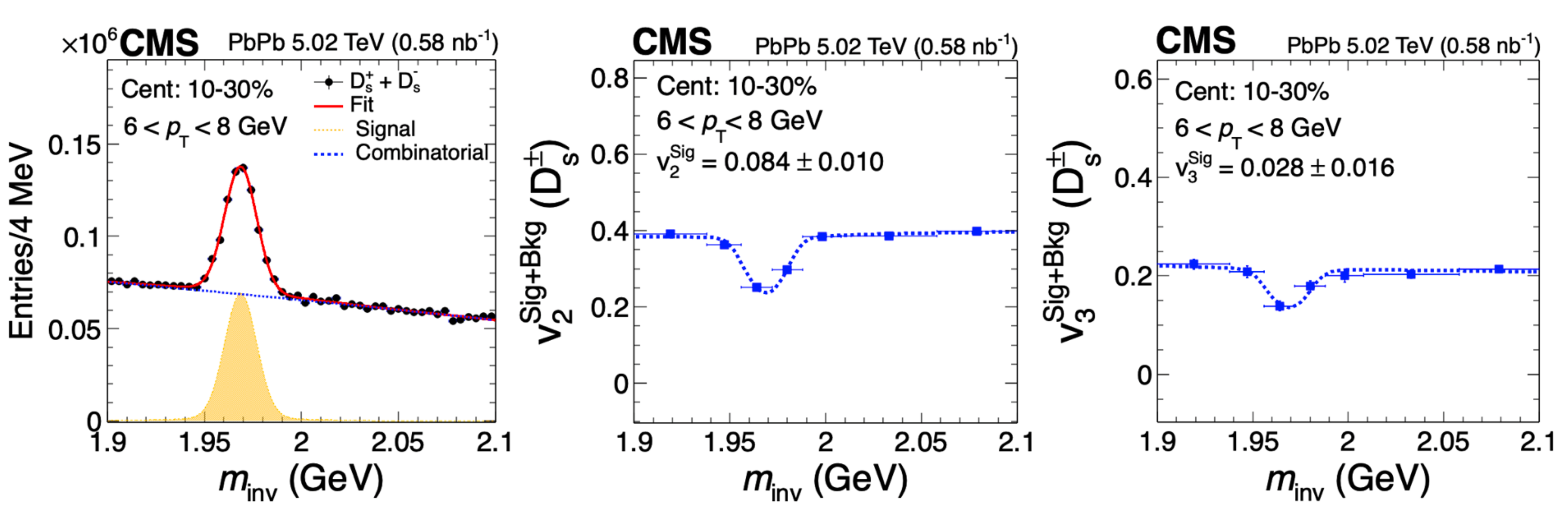}
    \topcaption{The left plot shows an example of the invariant mass ($\minv$) spectrum fit for \PDpms candidates in the \PDpms \pt interval 6--8\GeV for the centrality class 10--30\%. The middle and right plots show examples of the simultaneous fit where the $v_n$ value of the combined signal and background (${v_n}^\mathrm{Sig+Bkg}$) is modeled as a function of the invariant mass. The horizontal lines represent the \minv mass binning and the vertical lines (largely concealed by the corresponding markers) represent the statistical uncertainty.}
    \label{fig:vn_SF}
\end{figure*}

The \minv distribution for the \PDpms candidates includes a substantial combinatorial background, as seen in Fig.~\ref{fig:vn_SF} (left). This results in the $v_{n}$ values calculated using Eq.~(\ref{eq:scalar_product_vn}) having an \minv dependence, as shown in Fig.~\ref{fig:vn_SF} (middle and right) for the \vtwo and \vthree coefficients, respectively. To separate out the flow coefficients associated with the true signal $(v_n^\text{Sig})$ from those arising from the background $(v_n^\text{Bkg})$, a simultaneous fit is performed for the \minv distribution and the corresponding $v_n (\minv)$ distributions, using Eqs.~(\ref{eq:vn_signal}) and (\ref{eq:alpha}):
\begin{align}
\label{eq:vn_signal}
v_n^{\text{Sig} + \text{Bkg}}(\minv) &= \alpha(\minv)v_n^\text{Sig} + \left[1 - \alpha(\minv)\right]v_n^\text{Bkg}(\minv), \\
\intertext{with the factor}
\label{eq:alpha}
\alpha(\minv)  &= \frac{N^{\text{Sig}}(\minv)}{ N^{\text{Sig}}(\minv) + N^{\text{Bkg}}(\minv) }.
\end{align}
Here, ``$\text{Sig}$" refers to signal \PDpms candidates and ``$\text{Bkg}$" refers to the combinatorial background underlying the \PDpms  candidates~\cite{SP_ref_1}. The \minv distributions are fitted with two components, the sum of two Gaussian functions with the same mean but different widths for the \PDpms signal distribution, $N^{\text{Sig}}(\minv)$, and a second-order Chebyshev polynomial to model the combinatorial background distribution, $N^{\text{Bkg}}(\minv)$. The initial values of the widths and means of the $N^{\text{Sig}}(\minv)$ distribution used in the fit are taken from simulation studies. To allow for potential differences in the momentum resolution between simulation and data, the width parameters of the two Gaussian functions are multiplied by a common factor that is left free in the fit. In addition, the means are also allowed to be different in simulation and data. The $v_n^\text{Sig}$ coefficient is considered as the nominal value of $v_n$ in this analysis. The background $v_n^\text{Bkg}$ dependence on \minv is modeled with a linear function when performing the simultaneous fit to the \minv and $v_n^\mathrm{Sig + Bkg}(m_\mathrm{inv})$ distributions. Figure~\ref{fig:vn_SF} shows an example of the simultaneous fit to the \PDpms mass spectrum (left) and corresponding ${v_{2}}^\mathrm{Sig+Bkg}$ (middle) and ${v_{3}}^\mathrm{Sig+Bkg}$ (right) distributions. In this study, we focus on $v_2$ and $v_3$ as these are the primary harmonics that can be measured with meaningful precision. The higher-order harmonics are statistically inaccessible in the current dataset due to the combination of the limited \PDpms signal-to-background ratio and the rapid decrease in event-plane resolution for increasing harmonic orders, which significantly inflates the associated statistical uncertainties. The full analysis chain has been validated using Monte Carlo simulated data.

\section{Systematic uncertainties}
\label{sec:syst}

Systematic uncertainties are estimated by varying the conditions and procedures of the \vn determination. We considered seven sources of systematic uncertainty, as listed in Table~\ref{tab:summary_sys_v2_v3}. The systematic uncertainty associated with a given change in the analysis procedure is taken as the absolute value of the difference in the calculated \vn value before and after the change is applied. For each source, Table~\ref{tab:summary_sys_v2_v3} gives the range of associated systematic uncertainties when considering all \pt values and centrality classes. The calculated differences are found to be consistent across the measured \pt and centrality ranges, suggesting that the systematic estimates are not significantly biased by statistical fluctuations. Consequently, the Barlow test, often used to exclude statistically insignificant variations, was not performed, leading to a more conservative estimate of the total systematic uncertainty. The uncertainty associated with the signal fit function is found by comparing the results using a triple-Gaussian function to those obtained with the nominal two-Gaussian function. The uncertainty in the background underlying the \PDpms peak in the \minv spectrum is studied by changing the functional form of the combinatorial background from the nominal second-order Chebyshev polynomial dependence on \minv to a third-order polynomial dependence. The uncertainty arising from the background in the \vn distributions is found by comparing the results using a second-order polynomial rather than the nominal linear function to describe this background.

The systematic uncertainty for the \vn values resulting from the non-prompt contribution is determined by solving 
\begin{equation}
\label{eq:vn_signal_prompt}
v_n^{\text{Sig}}  = f_{\text{prompt}}v_n^{\text{prompt}} + \left(1 -  f_{\text{prompt}}\right)v_n^{\text{non-prompt}} 
\end{equation}
for two different cases: one, $v_{n,1}^{\text{Sig}}$, with the nominal selection corresponding to $\mathrm{DCA} < 0.01$ cm, and the other one, $v_{n,2}^{\text{Sig}}$, without having a DCA cut. Here, $f_{\text{prompt}}$ is the prompt fraction of \PDpms mesons, and $v_{n}^{\text{prompt}}$ and $v_{n}^{\text{non-prompt}}$ are the \vn values for the prompt and non-prompt \PDpms mesons, respectively. We estimate the $f_{\text{prompt}}$ for both cases with the help of a DCA template fit between data and simulation. Then solving for $v_n^{\text{prompt}}$ based on the two DCA cuts, we obtain 
\begin{equation}
\label{eq:vn_prompt}
v_n^{\text{prompt}}  =  v_{n,1}^{\text{Sig}}  +  \frac{1-f_{\text{prompt}, 1}}{f_{\text{prompt}, 1} - f_{\text{prompt}, 2}} \left(v_{n,1}^{\text{Sig}} -v_{n,2}^{\text{Sig}} \right),
\end{equation}
with the second term of this equation taken as the uncertainty resulting from the non-prompt contribution. A similar procedure is followed by the previous prompt \PDz analysis~\cite{Sirunyan:2020dvn}. 

The effect of the BDT selection is studied by varying the nominal BDT cuts by increments of $\pm$0.01, $\pm$0.02, $\pm$0.05, etc., and then using a linear function to extrapolate to the case where there is no BDT cut, as done in Ref.~\cite{lambdaC}. For the systematic study of acceptance and efficiency corrections related to \PDpms mesons, the product of the acceptance times the efficiency is increased (decreased) by 5\%, considered as a loose (tight) variation, to account for the residual differences between MC and data. The maximum variation in the \vn values between tight and loose cuts is considered as the respective uncertainty. Finally, the uncertainty related to the track quality is determined by tightening the tracking criteria to require $\chi^2/\text{dof}/N_\text{layers} < 0.15$. The total systematic uncertainty for each (\pt and centrality) bin is found by adding the individual contributions for that bin in quadrature.

\begin{table}[htbp!]
 \centering
 \topcaption{Summary of systematic uncertainties in the absolute differences between the nominal and the alternative analyses for the $\vtwo$ and $\vthree$ values.}
 \label{tab:summary_sys_v2_v3} 
 \cmsTable{
 \begin{tabular}{lcc}
\hline
 Source                       & \vtwo  & \vthree    \\
\hline
  Signal mass fit                         & 0 -- 0.009  & 0 -- 0.008 \\
  Bkg. mass fit                           & 0.001 -- 0.009  & 0 -- 0.006  \\
  Bkg. \vn fit                             & 0 -- 0.008  & 0 -- 0.006 \\
  Non-prompt contribution                 & 0 -- 0.012  & 0.001 -- 0.008 \\
  BDT selection                           & 0 -- 0.009  & 0 -- 0.009 \\
  Efficiency correction                   & 0 -- 0.019  & 0 -- 0.004 \\
  Tracking quality                      & 0 -- 0.008  & 0 -- 0.005 \\
  Total                                 &0.007 -- 0.025  &0.008 -- 0.013\\ 
  \hline                                
 \end{tabular}
 }
\end{table}

\section{Results}
\label{sec:res}

\begin{figure*}[ht!]
\centering
\includegraphics[width=0.99\textwidth]{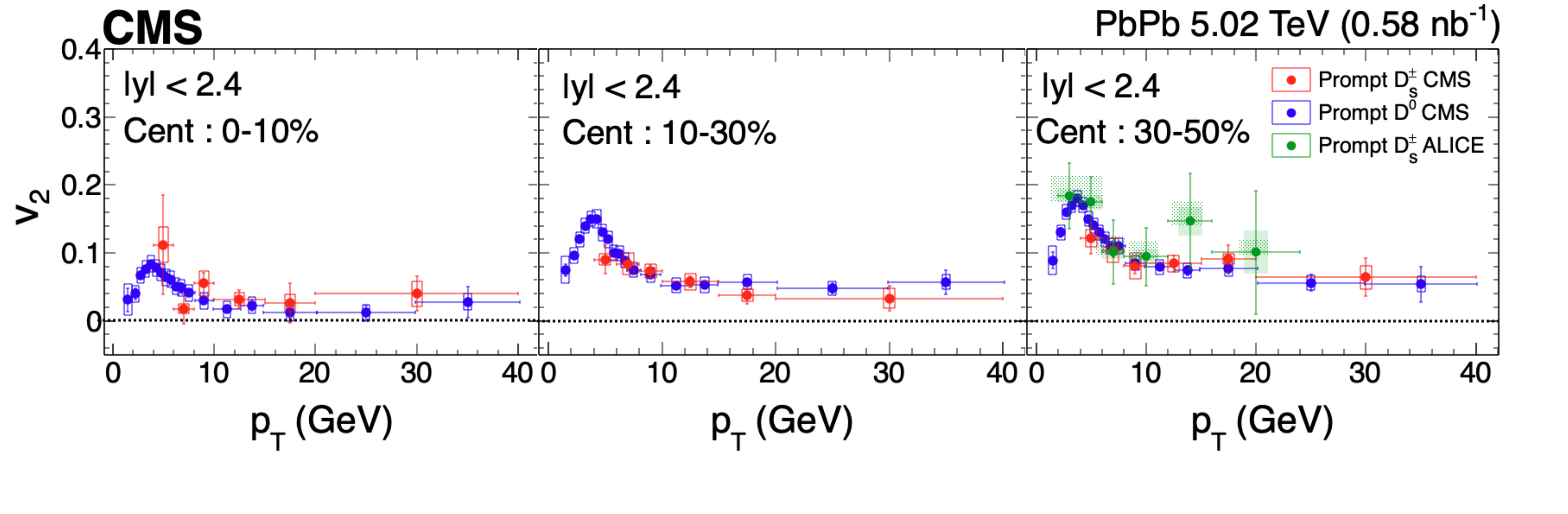}
\topcaption{The \vtwo coefficients for prompt \PDpms and prompt \PDz (from Ref.~\cite{Sirunyan:2020dvn}) mesons as a function of \PDpms \pt for the 0--10\% (left), 10--30\% (middle), and 30--50\% (right) centrality classes. The hatched area shown on top of the ALICE \vtwo results~\cite{Ds-Meson-ALICE} in the 30--50\% centrality class represents the contribution to the systematic uncertainty of the non-prompt contribution. The horizontal lines represent the \pt bin ranges. The vertical bars and the boxes represent statistical and systematic uncertainties, respectively.  }
\label{fig:v2_prompt}
\end{figure*}

\begin{figure*}[ht!]
\centering
\includegraphics[width=0.49\textwidth]{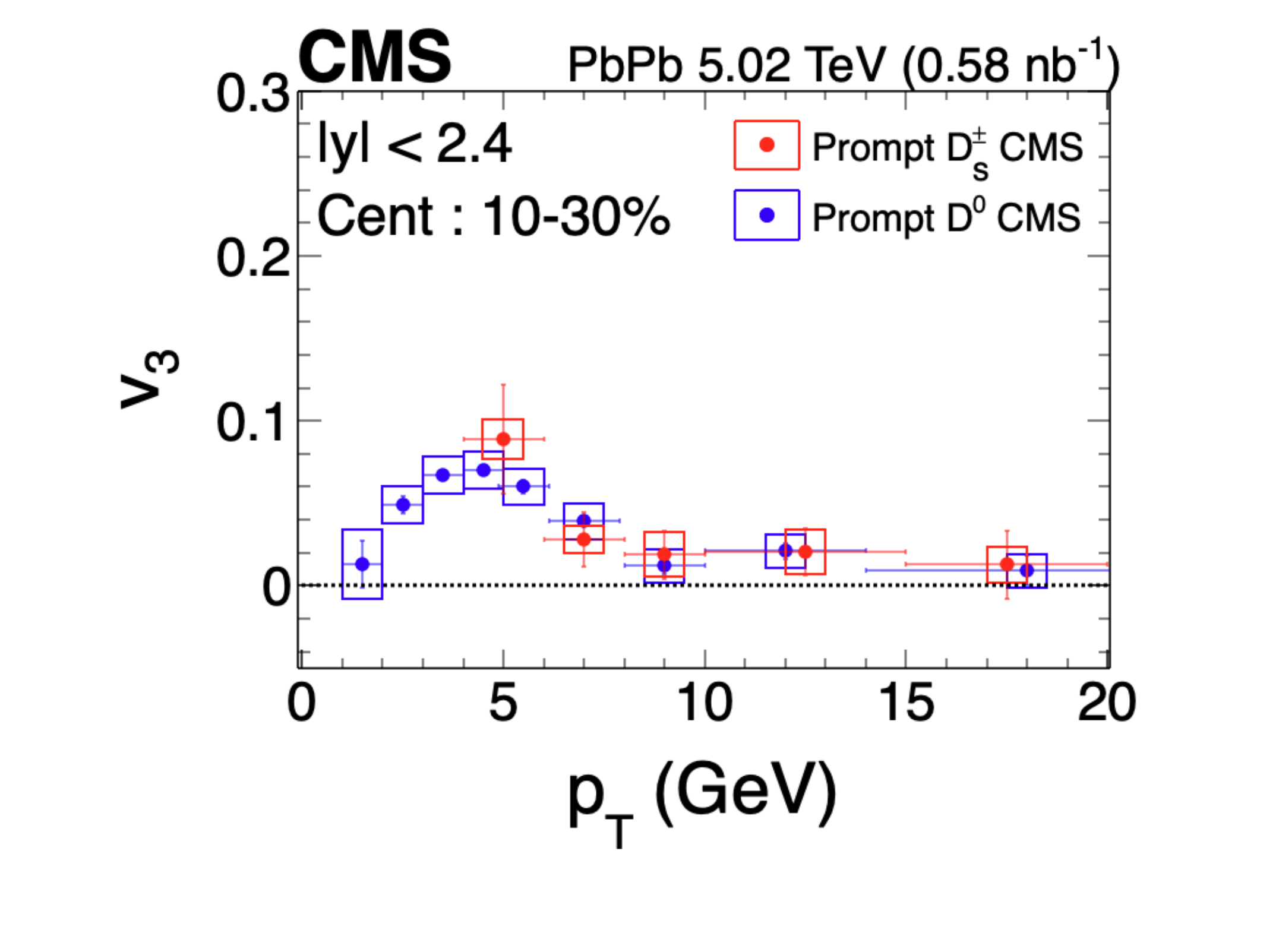}
\topcaption{The \vthree coefficients for prompt \PDpms and prompt \PDz (from Ref.~\cite{Sirunyan:2020dvn}) mesons as a function of \PDpms \pt for the 10--30\% centrality class. The horizontal lines represent the \pt bin ranges. The vertical bars and the boxes represent statistical and systematic uncertainties, respectively. }
\label{fig:v3_prompt}
\end{figure*}

The results for the prompt \PDpms meson \vtwo and \vthree flow harmonics are presented in Fig.~\ref{fig:v2_prompt} and Fig.~\ref{fig:v3_prompt}, along with the \vtwo results of the ALICE experiment for the 30--50\% centrality class~\cite{Ds-Meson-ALICE}. The figures also show the CMS \vtwo and \vthree results for \PDz mesons~\cite{Sirunyan:2020dvn} in the same centrality classes as measured here. The CMS \vtwo coefficients, averaged in the range $4 < \pt < 10\GeV$ are $\vtwo = 0.062 \pm 0.045 \stat \pm  0.018\syst$, $\vtwo = 0.083 \pm  0.014\stat \pm 0.012 \syst$, and $\vtwo = 0.101 \pm 0.016 \stat \pm 0.017 \syst $ in the centrality classes 0--10\%, 10--30\%, and 30--50\%, respectively. A positive \vtwo coefficient, with a significance of more than one standard deviation, is found in each of the measured centrality classes. The measured \vtwo coefficient peaks for \pt values in the range 4--6 \GeV and decreases with increasing \pt value, independently of the centrality class. The elliptic flow \vtwo values show a clear centrality dependence, with the largest values observed in the 30--50\% centrality class across all \pt intervals. The present measurement is consistent with that of ALICE~\cite{Ds-Meson-ALICE} in the centrality class 30--50\%, albeit with significantly higher precision and covering a wider kinematic region. Figure~\ref{fig:v3_prompt} shows the \PDpms \vthree results for 10--30\% central events in the range $4 < \pt < 10 \GeV$. Significant and positive values are observed. Unlike the elliptic flow \vtwo harmonic in this centrality class, which is dominated by the average lenticular geometry of the collision overlap region, the \vthree results are believed to arise primarily from event-by-event fluctuations in the initial-state density. Even with this different origin, the \PDpms and \PDz meson \vthree harmonics are very similar. The similar Fourier coefficients for the \PDpms and \PDz mesons suggest that the strangeness content of the \PDpms mesons does not have a significant influence on the hadronization processes that result from transit through a QGP medium in the measured \pt range.\\

\begin{figure*}[ht!]
\centering
\includegraphics[width=0.99\textwidth]{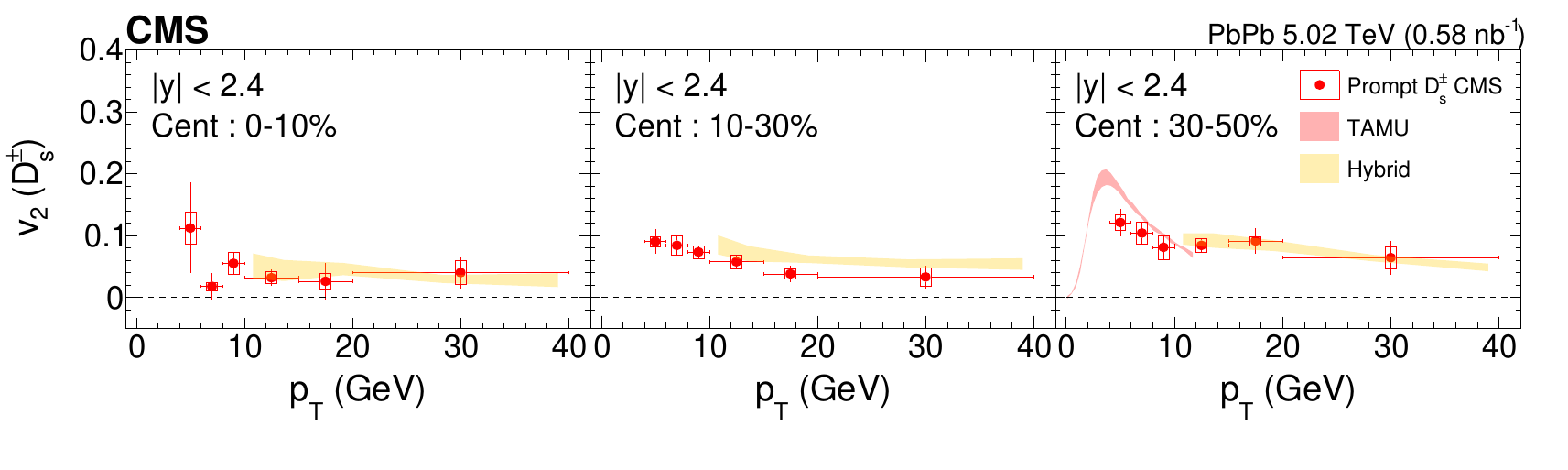}
\topcaption{The \vtwo coefficients for prompt \PDpms mesons as a function of \PDpms \pt for the 0--10\% (left), 10--30\% (middle), and 30--50\% (right) centrality classes. The horizontal lines represent the \pt bin ranges. The vertical bars and the boxes represent statistical and systematic uncertainties, respectively. The colored bands show theoretical predictions.}
\label{fig:v2_model}
\end{figure*}

To further investigate the physics mechanisms driving the \PDpms meson azimuthal anisotropy, in Fig.~\ref{fig:v2_model} the theoretical predictions from the TAMU~\cite{He:2014tamu} and HYBRID~\cite{hybrid-1, hybrid-2, hybrid-3} models, which provide theoretical uncertainty bands, are compared to the  measured \vtwo coefficients. The predictions for both models were obtained via private communication and calculated using the theoretical frameworks described in the cited references. The TAMU model computes the space-time evolution of the heavy-quark phase space distribution in the QGP using the Fokker-Planck equation implemented via Langevin dynamics, and assumes collisional energy loss without radiative contributions. For the 30--50\% centrality interval, the TAMU predictions are in qualitative agreement with the measured \vtwo results in the available range, although they indicate a more rapid rise as \pt values decrease from 10 \GeV to 4 \GeV. The HYBRID model combines perturbative QCD for jet evolution with insights from gauge-gravity duality to account for the strong-coupling effects of the QGP on parton energy loss. This allows the model to describe the jet quenching phenomenon by treating high-energy processes perturbatively and interactions with the medium through a strongly coupled framework. The HYBRID predictions yield good agreement with the data across all three centrality intervals in the available range of $\pt > 10 \GeV$.

\section{Summary}
\label{sec:sum}
The elliptic flow \vtwo and triangular flow \vthree Fourier coefficients for prompt \PDpms mesons in lead-lead (PbPb) collisions at a nucleon-nucleon center-of-mass energy of 5.02 \TeV have been measured using the CMS detector at the CERN LHC. The \vtwo results for transverse momenta in the range $4 < \pt < 40\GeV$ are presented in three ranges of collision centrality, corresponding to central, mid-central, and peripheral collisions. The \vthree results in the range $4<\pt<20\GeV$ are presented for mid-central collisions. The analysis employs the full rapidity coverage $\abs{y}<2.4$ of the CMS detector. The \vtwo results are measured with a precision that allows for a detailed comparison to \PDz mesons. The coefficients show a clear ordering by centrality and a pronounced \pt dependence in the 10--30\% centrality class. A non-zero \vthree coefficient is also seen for $4<\pt<10\GeV$, suggesting the influence of initial-state geometry fluctuations in \PDpms meson production. The \vtwo and \vthree coefficients of the \PDpms mesons, which contain a strange quark, are compared to those previously measured for non-strange \PDz mesons. The azimuthal anisotropies of \PDpms and \PDz mesons are found to be consistent within the measured \pt range. This result indicates that the anisotropies are dominated by the parent charm quark evolution in the QGP and that the underlying physics mechanisms, whether coalescence at low \pt or path-length dependent energy loss at high \pt, are not strongly sensitive to the flavor of the light quark involved in hadronization. Predictions from the TAMU and HYBRID theoretical models are compared to the measured \vtwo coefficients. Both models provide a good description of the experimental data within uncertainties and the available \pt range, supporting the conclusion that the observed \PDpms meson azimuthal anisotropy is strongly driven by the interactions between charm quarks and the expanding QGP medium.

\begin{acknowledgments}
We congratulate our colleagues in the CERN accelerator departments for the excellent performance of the LHC and thank the technical and administrative staffs at CERN and at other CMS institutes for their contributions to the success of the CMS effort. In addition, we gratefully acknowledge the computing centers and personnel of the Worldwide LHC Computing Grid and other centers for delivering so effectively the computing infrastructure essential to our analyses. Finally, we acknowledge the enduring support for the construction and operation of the LHC, the CMS detector, and the supporting computing infrastructure provided by the following funding agencies: SC (Armenia), BMBWF and FWF (Austria); FNRS and FWO (Belgium); CNPq, CAPES, FAPERJ, FAPERGS, and FAPESP (Brazil); MES and BNSF (Bulgaria); CERN; CAS, MoST, and NSFC (China); MINCIENCIAS (Colombia); MSES and CSF (Croatia); RIF (Cyprus); SENESCYT (Ecuador); ERC PRG and PSG, TARISTU24-TK10 and MoER TK202 (Estonia); Academy of Finland, MEC, and HIP (Finland); CEA and CNRS/IN2P3 (France); SRNSF (Georgia); BMFTR, DFG, and HGF (Germany); GSRI (Greece); NKFIH (Hungary); DAE and DST (India); IPM (Iran); SFI (Ireland); INFN (Italy); MSIT and NRF (Republic of Korea); MES (Latvia); LMTLT (Lithuania); MOE and UM (Malaysia); BUAP, CINVESTAV, CONACYT, LNS, SEP, and UASLP-FAI (Mexico); MOS (Montenegro); MBIE (New Zealand); PAEC (Pakistan); MES, NSC, and NAWA (Poland); FCT (Portugal); MESTD (Serbia); MICIU/AEI and PCTI (Spain); MOSTR (Sri Lanka); Swiss Funding Agencies (Switzerland); MST (Taipei); MHESI (Thailand); TUBITAK and TENMAK (T\"{u}rkiye); NASU (Ukraine); STFC (United Kingdom); DOE and NSF (USA).

\hyphenation{Rachada-pisek} Individuals have received support from the Marie-Curie programme and the European Research Council and Horizon 2020 Grant, contract Nos.\ 675440, 724704, 752730, 758316, 765710, 824093, 101115353, 101002207, 101001205, and COST Action CA16108 (European Union); the Leventis Foundation; the Alfred P.\ Sloan Foundation; the Alexander von Humboldt Foundation; the Science Committee, project no. 22rl-037 (Armenia); the Fonds pour la Formation \`a la Recherche dans l'Industrie et dans l'Agriculture (FRIA) and Fonds voor Wetenschappelijk Onderzoek contract No. 1228724N (Belgium); the Beijing Municipal Science \& Technology Commission, No. Z191100007219010, the Fundamental Research Funds for the Central Universities, the Ministry of Science and Technology of China under Grant No. 2023YFA1605804, the Natural Science Foundation of China under Grant No. 12535004, and USTC Research Funds of the Double First-Class Initiative No.\ YD2030002017 (China); the Ministry of Education, Youth and Sports (MEYS) of the Czech Republic; the Shota Rustaveli National Science Foundation, grant FR-22-985 (Georgia); the Deutsche Forschungsgemeinschaft (DFG), among others, under Germany's Excellence Strategy -- EXC 2121 ``Quantum Universe" -- 390833306, and under project number 400140256 - GRK2497; the Hellenic Foundation for Research and Innovation (HFRI), Project Number 2288 (Greece); the Hungarian Academy of Sciences, the New National Excellence Program - \'UNKP, the NKFIH research grants K 131991, K 133046, K 138136, K 143460, K 143477, K 146913, K 146914, K 147048, 2020-2.2.1-ED-2021-00181, TKP2021-NKTA-64, and 2025-1.1.5-NEMZ\_KI-2025-00004 (Hungary); the Council of Science and Industrial Research, India; ICSC -- National Research Center for High Performance Computing, Big Data and Quantum Computing, FAIR -- Future Artificial Intelligence Research, and CUP I53D23001070006 (Mission 4 Component 1), funded by the NextGenerationEU program (Italy); the Latvian Council of Science; the Ministry of Education and Science, project no. 2022/WK/14, and the National Science Center, contracts Opus 2021/41/B/ST2/01369, 2021/43/B/ST2/01552, 2023/49/B/ST2/03273, and the NAWA contract BPN/PPO/2021/1/00011 (Poland); the Funda\c{c}\~ao para a Ci\^encia e a Tecnologia (Portugal); the National Priorities Research Program by Qatar National Research Fund; MICIU/AEI/10.13039/501100011033, ERDF/EU, ``European Union NextGenerationEU/PRTR", and Programa Severo Ochoa del Principado de Asturias (Spain); the Chulalongkorn Academic into Its 2nd Century Project Advancement Project, the National Science, Research and Innovation Fund program IND\_FF\_68\_369\_2300\_097, and the Program Management Unit for Human Resources \& Institutional Development, Research and Innovation, grant B39G680009 (Thailand); the Eric \& Wendy Schmidt Fund for Strategic Innovation through the CERN Next Generation Triggers project under grant agreement number SIF-2023-004; the Kavli Foundation; the Nvidia Corporation; the SuperMicro Corporation; the Welch Foundation, contract C-1845; and the Weston Havens Foundation (USA).
  
\end{acknowledgments}

\bibliography{auto_generated}

\providecommand{\href}[2]{#2}\begingroup\raggedright\begin{thebibliography}{10}%
\makeatletter
\providecommand{\hrefCMSnoop }[0]{\@secondoftwo}%
\makeatother
\providecommand{\doi}{\texttt{doi:}\begingroup \urlstyle{tt}\Url}

\bibitem{BRAHAMS:2005}
\hrefCMSnoop {}{{BRAHMS} Collaboration, ``{Quark-gluon plasma and color glass
  condensate at RHIC? The perspective from the BRAHMS experiment}'',} \textit{
  Nucl. Phys. A} \textbf{ 757} (2005) 1,
  \href{http://dx.doi.org/10.1016/j.nuclphysa.2005.02.130}{\doi{10.1016/j.nuclphysa.2005.02.130}},
\href{http://www.arXiv.org/abs/nucl-ex/0410020}{\texttt{arXiv:nucl-ex/0410020}}.

\bibitem{PHOBOS:2005}
\hrefCMSnoop {}{{PHOBOS} Collaboration, ``{The PHOBOS perspective on
  discoveries at RHIC}'',} \textit{ Nucl. Phys. A} \textbf{ 757} (2005) 28,
  \href{http://dx.doi.org/10.1016/j.nuclphysa.2005.03.084}{\doi{10.1016/j.nuclphysa.2005.03.084}},
\href{http://www.arXiv.org/abs/nucl-ex/0410022}{\texttt{arXiv:nucl-ex/0410022}}.

\bibitem{STAR:2005}
\hrefCMSnoop {}{{STAR} Collaboration, ``{Experimental and theoretical
  challenges in the search for the quark-gluon plasma: the STAR Collaboration's
  critical assessment of the evidence from RHIC collisions}'',} \textit{ Nucl.
  Phys. A} \textbf{ 757} (2005) 102,
  \href{http://dx.doi.org/10.1016/j.nuclphysa.2005.03.085}{\doi{10.1016/j.nuclphysa.2005.03.085}},
\href{http://www.arXiv.org/abs/nucl-ex/0501009}{\texttt{arXiv:nucl-ex/0501009}}.

\bibitem{PHENIX:2005}
\hrefCMSnoop {}{{PHENIX} Collaboration, ``{Formation of dense partonic matter
  in relativistic nucleus-nucleus collisions at RHIC: experimental evaluation
  by the PHENIX Collaboration}'',} \textit{ Nucl. Phys. A} \textbf{ 757} (2005)
  184,
  \href{http://dx.doi.org/10.1016/j.nuclphysa.2005.03.086}{\doi{10.1016/j.nuclphysa.2005.03.086}},
\href{http://www.arXiv.org/abs/nucl-ex/0410003}{\texttt{arXiv:nucl-ex/0410003}}.

\bibitem{cms2025overview}
\hrefCMSnoop {}{{CMS Collaboration}, ``Overview of high-density {QCD} studies
  with the {CMS} experiment at the {LHC}'',} \textit{ Phys. Repts.} \textbf{
  1115} (2025) 219,
  \href{http://dx.doi.org/10.1016/j.physrep.2024.11.007}{\doi{10.1016/j.physrep.2024.11.007}},
  \href{http://www.arXiv.org/abs/2405.10785}{\texttt{arXiv:2405.10785}}.

\bibitem{QGP_ref_1}
\hrefCMSnoop {}{P.~Braun-Munzinger, V.~Koch, T.~Sch{\"a}fer, and J.~Stachel,
  ``Properties of hot and dense matter from relativistic heavy ion
  collisions'',} \textit{ Phys. Repts.} \textbf{ 621} (2016) 76,
  \href{http://dx.doi.org/10.1016/j.physrep.2015.12.003}{\doi{10.1016/j.physrep.2015.12.003}},
  \href{http://www.arXiv.org/abs/1510.00442}{\texttt{arXiv:1510.00442}}.

\bibitem{QGP_ref_2}
\hrefCMSnoop {}{W.~Busza, K.~Rajagopal, and W.~Van Der~Schee, ``Heavy ion
  collisions: the big picture and the big questions'',} \textit{ Ann. Rev.
  Nucl. Part. Sci.} \textbf{ 68} (2018) 339,
  \href{http://dx.doi.org/10.1146/annurev-nucl-101917-020852}{\doi{10.1146/annurev-nucl-101917-020852}},
  \href{http://www.arXiv.org/abs/1802.04801}{\texttt{arXiv:1802.04801}}.

\bibitem{Ollitrault:1992bk}
\hrefCMSnoop {}{J.-Y. Ollitrault, ``{Anisotropy as a signature of transverse
  collective flow}'',} \textit{ Phys. Rev. D} \textbf{ 46} (1992) 229,
\href{http://dx.doi.org/10.1103/PhysRevD.46.229}{\doi{10.1103/PhysRevD.46.229}}.

\bibitem{Heinz:2013th}
\hrefCMSnoop {}{U.~Heinz and R.~Snellings, ``{Collective flow and viscosity in
  relativistic heavy-ion collisions}'',} \textit{ Ann. Rev. Nucl. Part. Sci.}
  \textbf{ 63} (2013) 123,
  \href{http://dx.doi.org/10.1146/annurev-nucl-102212-170540}{\doi{10.1146/annurev-nucl-102212-170540}},
\href{http://www.arXiv.org/abs/1301.2826}{\texttt{arXiv:1301.2826}}.

\bibitem{Gale:2013da}
\hrefCMSnoop {}{C.~Gale, S.~Jeon, and B.~Schenke, ``{Hydrodynamic modeling of
  heavy-ion collisions}'',} \textit{ Int. J. Mod. Phys. A} \textbf{ 28} (2013)
  1340011,
  \href{http://dx.doi.org/10.1142/S0217751X13400113}{\doi{10.1142/S0217751X13400113}},
\href{http://www.arXiv.org/abs/1301.5893}{\texttt{arXiv:1301.5893}}.

\bibitem{Alver:2010prc}
\hrefCMSnoop {}{B.~Alver and G.~Roland, ``{Collision-geometry fluctuations and
  triangular flow in heavy-ion collisions}'',} \textit{ Phys. Rev. C} \textbf{
  81} (2010) 054905,
  \href{http://dx.doi.org/10.1103/PhysRevC.81.054905}{\doi{10.1103/PhysRevC.81.054905}},
\href{http://www.arXiv.org/abs/1003.0194}{\texttt{arXiv:1003.0194}}.

\bibitem{Prino:2016hf}
\hrefCMSnoop {}{F.~Prino and R.~Rapp, ``{Open heavy flavor in QCD matter and in
  nuclear collisions}'',} \textit{ J. Phys. G} \textbf{ 43} (2016) 093002,
  \href{http://dx.doi.org/10.1088/0954-3899/43/9/093002}{\doi{10.1088/0954-3899/43/9/093002}},
\href{http://www.arXiv.org/abs/1603.00529}{\texttt{arXiv:1603.00529}}.

\bibitem{Beraudo:2018enl}
\hrefCMSnoop {}{R.~Rapp { et~al.}, ``Extraction of heavy-flavor transport
  coefficients in {QCD} matter'',} \textit{ Nucl. Phys. A} \textbf{ 979} (2018)
  21,
  \href{http://dx.doi.org/10.1016/j.nuclphysa.2018.09.002}{\doi{10.1016/j.nuclphysa.2018.09.002}},
\href{http://www.arXiv.org/abs/1803.03824}{\texttt{arXiv:1803.03824}}.

\bibitem{vanHes:2006therm}
\hrefCMSnoop {}{H.~van Hees, V.~Greco, and R.~Rapp, ``Heavy-quark probes of the
  quark-gluon plasma and interpretation of recent data taken at the {BNL}
  {R}elativistic {H}eavy {I}on {C}ollider'',} \textit{ Phys. Rev. C} \textbf{
  73} (2006) 034913,
  \href{http://dx.doi.org/10.1103/PhysRevC.73.034913}{\doi{10.1103/PhysRevC.73.034913}},
  \href{http://www.arXiv.org/abs/nucl-th/0508055}{\texttt{arXiv:nucl-th/0508055}}.

\bibitem{Gyulassy2001hptv2}
\hrefCMSnoop {}{M.~Gyulassy, I.~Vitev, and X.-N. Wang, ``High {\pt} azimuthal
  asymmetry in noncentral {A+A} at {RHIC}'',} \textit{ Phys. Rev. Lett.}
  \textbf{ 86} (2001) 2537,
  \href{http://dx.doi.org/10.1103/PhysRevLett.86.2537}{\doi{10.1103/PhysRevLett.86.2537}},
\href{http://www.arXiv.org/abs/nucl-th/0012092}{\texttt{arXiv:nucl-th/0012092}}.

\bibitem{Shuryak2002hptv2}
\hrefCMSnoop {}{E.~V. Shuryak, ``Azimuthal asymmetry at large {\pt} seem to be
  too large for a pure 'jet quenching''',} \textit{ Phys. Rev. C} \textbf{ 66}
  (2002) 027902,
  \href{http://dx.doi.org/10.1103/PhysRevC.66.027902}{\doi{10.1103/PhysRevC.66.027902}},
\href{http://www.arXiv.org/abs/nucl-th/0112042}{\texttt{arXiv:nucl-th/0112042}}.

\bibitem{Acharya:2018dvn}
\hrefCMSnoop {}{{ALICE Collaboration}, ``{D-meson azimuthal anisotropy in
  midcentral Pb-Pb collisions at $\sqrtsNN = 5.02\TeV$}'',} \textit{ Phys. Rev.
  Lett.} \textbf{ 120} (2018) 102301,
  \href{http://dx.doi.org/10.1103/PhysRevLett.120.102301}{\doi{10.1103/PhysRevLett.120.102301}},
\href{http://www.arXiv.org/abs/1707.01005}{\texttt{arXiv:1707.01005}}.

\bibitem{Sirunyan:2017plt}
\hrefCMSnoop {}{{CMS Collaboration}, ``Measurement of prompt {$\PDz$} meson
  azimuthal anisotropy in {Pb-Pb} collisions at {$\sqrtsNN=5.02\TeV$}'',}
  \textit{ Phys. Rev. Lett.} \textbf{ 120} (2018) 202301,
  \href{http://dx.doi.org/10.1103/PhysRevLett.120.202301}{\doi{10.1103/PhysRevLett.120.202301}},
\href{http://www.arXiv.org/abs/1708.03497}{\texttt{arXiv:1708.03497}}.

\bibitem{Sirunyan:2020dvn}
\hrefCMSnoop {}{{CMS Collaboration}, ``{Measurement of prompt $\mathrm{D}^0$
  and $\overline{\mathrm{D}}^0$ meson azimuthal anisotropy and search for
  strong electric fields in PbPb collisions at $\sqrtsNN= 5.02$ TeV}'',}
  \textit{ Phys. Lett. B} \textbf{ 816} (2021) 136253,
  \href{http://dx.doi.org/10.1016/j.physletb.2021.136253}{\doi{10.1016/j.physletb.2021.136253}},
\href{http://www.arXiv.org/abs/2009.12628}{\texttt{arXiv:2009.12628}}.

\bibitem{Aad:2020bvn}
\hrefCMSnoop {}{{ATLAS Collaboration}, ``{Measurement of azimuthal anisotropy
  of muons from charm and bottom hadrons in Pb+Pb collisions at $\sqrtsNN =
  5.02\TeV$ with the ATLAS detector}'',} \textit{ Phys. Lett. B} \textbf{ 807}
  (2020) 135595,
  \href{http://dx.doi.org/10.1016/j.physletb.2020.135595}{\doi{10.1016/j.physletb.2020.135595}},
\href{http://www.arXiv.org/abs/2003.03565}{\texttt{arXiv:2003.03565}}.

\bibitem{Acharya:2020jpsivn}
\hrefCMSnoop {}{{ALICE Collaboration}, ``{\cPJgy} elliptic and triangular flow
  in {Pb-Pb} collisions at {$\sqrtsNN = 5.02\TeV$}'',} \textit{ JHEP} \textbf{
  10} (2020) 141,
  \href{http://dx.doi.org/10.1007/JHEP10(2020)141}{\doi{10.1007/JHEP10(2020)141}},
\href{http://www.arXiv.org/abs/2005.14518}{\texttt{arXiv:2005.14518}}.

\bibitem{Acharya:2021Dvn}
\hrefCMSnoop {}{{ALICE Collaboration}, ``{Transverse-momentum and event-shape
  dependence of D-meson flow harmonics in Pb-Pb collisions at $\sqrtsNN =
  5.02\TeV$}'',} \textit{ Phys. Lett. B} \textbf{ 813} (2021) 136054,
  \href{http://dx.doi.org/10.1016/j.physletb.2020.136054}{\doi{10.1016/j.physletb.2020.136054}},
\href{http://www.arXiv.org/abs/2005.11131}{\texttt{arXiv:2005.11131}}.

\bibitem{Sirunyan:2022dv24}
\hrefCMSnoop {}{{CMS Collaboration}, ``Probing charm quark dynamics via
  multiparticle correlations in {\PbPb} collisions at {$\sqrtsNN = 5.02
  \TeV$}'',} \textit{ Phys. Rev. Lett.} \textbf{ 129} (2022) 022001,
  \href{http://dx.doi.org/10.1103/PhysRevLett.129.022001}{\doi{10.1103/PhysRevLett.129.022001}},
\href{http://www.arXiv.org/abs/2112.12236}{\texttt{arXiv:2112.12236}}.

\bibitem{tumasyan2023measurements}
\hrefCMSnoop {}{{CMS Collaboration}, ``Measurements of the azimuthal anisotropy
  of prompt and non-prompt charmonia in {PbPb} collisions at 5.02 {TeV}'',}
  \textit{ JHEP} \textbf{ 10} (2023) 115,
  \href{http://dx.doi.org/10.1007/jhep10(2023)115}{\doi{10.1007/jhep10(2023)115}}.

\bibitem{strangeQGP-1}
\hrefCMSnoop {}{P.~Koch, B.~M{\"u}ller, and J.~Rafelski, ``Strangeness in
  relativistic heavy ion collisions'',} \textit{ Phys. Repts.} \textbf{ 142}
  (1986) 167,
  \href{http://dx.doi.org/10.1016/0370-1573(86)90096-7}{\doi{10.1016/0370-1573(86)90096-7}}.

\bibitem{strangeQGP-2}
\hrefCMSnoop {}{{ALICE Collaboration}, ``Multi-strange baryon production at
  mid-rapidity in {PbPb} collisions at 2.76 tev'',} \textit{ Phys. Lett. B}
  \textbf{ 728} (2014) 216,
  \href{http://dx.doi.org/10.1016/j.physletb.2013.11.048}{\doi{10.1016/j.physletb.2013.11.048}}.

\bibitem{rapp2005}
\hrefCMSnoop {}{R.~Rapp, ``Quark coalescence and charm(onium) in {QGP}'',}
  \textit{ Eur. Phys. J. C} \textbf{ 43} (2005) 91,
  \href{http://dx.doi.org/10.1140/epjc/s2005-02215-0}{\doi{10.1140/epjc/s2005-02215-0}},
  \href{http://www.arXiv.org/abs/hep-ph/0502208}{\texttt{arXiv:hep-ph/0502208}}.

\bibitem{he2012}
\hrefCMSnoop {}{M.~He, R.~J. Fries, and R.~Rapp, ``Heavy-quark diffusion and
  hadronization in quark-gluon plasma'',} \textit{ Phys. Rev. C} \textbf{ 86}
  (2012) 014903,
  \href{http://dx.doi.org/10.1103/PhysRevC.86.014903}{\doi{10.1103/PhysRevC.86.014903}},
  \href{http://www.arXiv.org/abs/1106.6006}{\texttt{arXiv:1106.6006}}.

\bibitem{PhysRevLett.123.192301}
\hrefCMSnoop {}{{ALICE Collaboration}, ``Measurement of {$\Upsilon(1S)$}
  elliptic flow at forward rapidity in {Pb-Pb} collisions at
  {$\sqrtsNN=5.02\TeV$}'',} \textit{ Phys. Rev. Lett.} \textbf{ 123} (2019)
  192301,
  \href{http://dx.doi.org/10.1103/PhysRevLett.123.192301}{\doi{10.1103/PhysRevLett.123.192301}},
  \href{http://www.arXiv.org/abs/1907.03169}{\texttt{arXiv:1907.03169}}.

\bibitem{HEPData}
\hrefCMSnoop {}{{CMS Collaboration}, ``{HEP} data record for this analysis'',}
  2025.
\newblock
  \href{http://dx.doi.org/10.17182/hepdata.158600}{\doi{10.17182/hepdata.158600}}.

\bibitem{Chatrchyan:2008zzk}
\hrefCMSnoop {}{{CMS Collaboration}, ``Development of the {CMS} detector for
  the {CERN} {LHC} {Run} 3'',} \textit{ JINST} \textbf{ 19} (2024) P05064,
  \href{http://dx.doi.org/10.1088/1748-0221/19/05/P05064}{\doi{10.1088/1748-0221/19/05/P05064}}.

\bibitem{CMS:2020cmk}
\hrefCMSnoop {}{{CMS Collaboration}, ``{Performance of the {CMS} level-1
  trigger in proton-proton collisions at $\sqrt{s} = 13$\,TeV}'',} \textit{
  JINST} \textbf{ 15} (2020) P10017,
  \href{http://dx.doi.org/10.1088/1748-0221/15/10/P10017}{\doi{10.1088/1748-0221/15/10/P10017}},
  \href{http://www.arXiv.org/abs/2006.10165}{\texttt{arXiv:2006.10165}}.

\bibitem{CMS:2016ngn}
\hrefCMSnoop {}{{CMS Collaboration}, ``Performance of the {CMS} high-level
  trigger during {LHC} {Run} 2'',} \textit{ JINST} \textbf{ 19} (2024) P11021,
  \href{http://dx.doi.org/10.1088/1748-0221/19/11/P11021}{\doi{10.1088/1748-0221/19/11/P11021}}.

\bibitem{CMS:2020uim}
\hrefCMSnoop {}{{CMS Collaboration}, ``Electron and photon reconstruction and
  identification with the {CMS} experiment at the {CERN} {LHC}'',} \textit{
  JINST} \textbf{ 16} (2021) P05014,
  \href{http://dx.doi.org/10.1088/1748-0221/16/05/P05014}{\doi{10.1088/1748-0221/16/05/P05014}},
  \href{http://www.arXiv.org/abs/2012.06888}{\texttt{arXiv:2012.06888}}.

\bibitem{CMS:2018rym}
\hrefCMSnoop {}{{CMS Collaboration}, ``{Performance of the CMS muon detector
  and muon reconstruction with proton-proton collisions at $\sqrt{s}=$ 13
  TeV}'',} \textit{ JINST} \textbf{ 13} (2018) P06015,
  \href{http://dx.doi.org/10.1088/1748-0221/13/06/P06015}{\doi{10.1088/1748-0221/13/06/P06015}},
  \href{http://www.arXiv.org/abs/1804.04528}{\texttt{arXiv:1804.04528}}.

\bibitem{CMS:2014pgm}
\hrefCMSnoop {}{{CMS Collaboration}, ``{Description and performance of track
  and primary-vertex reconstruction with the CMS tracker}'',} \textit{ JINST}
  \textbf{ 9} (2014) P10009,
  \href{http://dx.doi.org/10.1088/1748-0221/9/10/P10009}{\doi{10.1088/1748-0221/9/10/P10009}},
  \href{http://www.arXiv.org/abs/1405.6569}{\texttt{arXiv:1405.6569}}.

\bibitem{CMS:2021xjt}
\hrefCMSnoop {}{{CMS Collaboration}, ``Precision luminosity measurement in
  proton-proton collisions at {$\sqrt{s}=13\TeV$} in 2015 and 2016 at {CMS}'',}
  \textit{ Eur. Phys. J. C} \textbf{ 81} (2021) 800,
  \href{http://dx.doi.org/10.1140/epjc/s10052-021-09538-2}{\doi{10.1140/epjc/s10052-021-09538-2}},
  \href{http://www.arXiv.org/abs/2104.01927}{\texttt{arXiv:2104.01927}}.

\bibitem{cmsPbPbLumi5p02TeV}
\hrefCMSnoop {}{{CMS Collaboration}, ``Luminosity measurement for lead-lead
  collisions at {$\sqrtsNN=5.02\TeV$} in 2015 and 2018 at {CMS}'',} 2025.
  \href{http://www.arXiv.org/abs/2503.03946}{\texttt{arXiv:2503.03946}}.
  Submitted to {EPJC}.

\bibitem{upc}
\hrefCMSnoop {}{A.~J. Baltz { et~al.}, ``{The Physics of Ultraperipheral
  Collisions at the {LHC}}'',} \textit{ Phys. Rept.} \textbf{ 458} (2008) 1,
  \href{http://dx.doi.org/10.1016/j.physrep.2007.12.001}{\doi{10.1016/j.physrep.2007.12.001}},
  \href{http://www.arXiv.org/abs/0706.3356}{\texttt{arXiv:0706.3356}}.

\bibitem{CMS:2011iwn}
\hrefCMSnoop {}{{CMS Collaboration}, ``Observation and studies of jet quenching
  in {\PbPb} collisions at {$\sqrtsNN=2.76\TeV$}'',} \textit{ Phys. Rev. C}
  \textbf{ 84} (2011) 024906,
  \href{http://dx.doi.org/10.1103/PhysRevC.84.024906}{\doi{10.1103/PhysRevC.84.024906}},
  \href{http://www.arXiv.org/abs/1102.1957}{\texttt{arXiv:1102.1957}}.

\bibitem{Sjostrand:2014zea}
T.~Sj{\"o}strand\hrefCMSnoop {}{ { et~al.}, ``An introduction to {\PYTHIA}
  8.2'',} \textit{ Comput. Phys. Commun.} \textbf{ 191} (2015) 159,
  \href{http://dx.doi.org/10.1016/j.cpc.2015.01.024}{\doi{10.1016/j.cpc.2015.01.024}},
\href{http://www.arXiv.org/abs/1410.3012}{\texttt{arXiv:1410.3012}}.

\bibitem{Sirunyan:2019dfx}
\hrefCMSnoop {}{{CMS Collaboration}, ``Extraction and validation of a new set
  of {CMS} {\PYTHIA}8 tunes from underlying-event measurements'',} \textit{
  Eur. Phys. J. C} \textbf{ 80} (2020) 4,
  \href{http://dx.doi.org/10.1140/epjc/s10052-019-7499-4}{\doi{10.1140/epjc/s10052-019-7499-4}},
  \href{http://www.arXiv.org/abs/1903.12179}{\texttt{arXiv:1903.12179}}.

\bibitem{Lange:2001uf}
\hrefCMSnoop {}{D.~J. Lange, ``The {\EVTGEN} particle decay simulation
  package'',} \textit{ Nucl. Instrum. Meth. A} \textbf{ 462} (2001) 152,
  \href{http://dx.doi.org/10.1016/S0168-9002(01)00089-4}{\doi{10.1016/S0168-9002(01)00089-4}}.

\bibitem{Lokhtin:2005px}
\hrefCMSnoop {}{I.~P. Lokhtin and A.~M. Snigirev, ``A model of jet quenching in
  ultrarelativistic heavy ion collisions and high-$\pt$ hadron spectra at
  {RHIC}'',} \textit{ Eur. Phys. J. C} \textbf{ 45} (2006) 211,
  \href{http://dx.doi.org/10.1140/epjc/s2005-02426-3}{\doi{10.1140/epjc/s2005-02426-3}},
\href{http://www.arXiv.org/abs/hep-ph/0506189}{\texttt{arXiv:hep-ph/0506189}}.

\bibitem{Khachatryan:2016odn}
\hrefCMSnoop {}{{CMS Collaboration}, ``Charged-particle nuclear modification
  factors in {PbPb} and {pPb} collisions at {$\sqrtsNN=5.02\TeV$}'',} \textit{
  JHEP} \textbf{ 04} (2017) 039,
  \href{http://dx.doi.org/10.1007/JHEP04(2017)039}{\doi{10.1007/JHEP04(2017)039}},
\href{http://www.arXiv.org/abs/1611.01664}{\texttt{arXiv:1611.01664}}.

\bibitem{Agostinelli:2002hh}
\hrefCMSnoop {}{{GEANT4} Collaboration, ``{\GEANTfour}---a simulation
  toolkit'',} \textit{ Nucl. Instrum. Meth. A} \textbf{ 506} (2003) 250,
\href{http://dx.doi.org/10.1016/S0168-9002(03)01368-8}{\doi{10.1016/S0168-9002(03)01368-8}}.

\bibitem{Workman:2024pdg}
\hrefCMSnoop {}{{Particle Data Group}, R.~L. Workman { et~al.}, ``Review of
  particle physics'',} \textit{ Phys. Rev. D} \textbf{ 110} (2024) 030001,
  \href{http://dx.doi.org/10.1103/PhysRevD.110.030001}{\doi{10.1103/PhysRevD.110.030001}}.

\bibitem{Hocker:2007ht}
\hrefCMSnoop {}{A.~Hocker { et~al.}, ``{TMVA} --- toolkit for multivariate data
  analysis'',} \textit{ PoS} \textbf{ ACAT} (2007) 040,
  \href{http://dx.doi.org/10.48550/arXiv.physics/0703039}{\doi{10.48550/arXiv.physics/0703039}},
  \href{http://www.arXiv.org/abs/physics/0703039}{\texttt{arXiv:physics/0703039}}.

\bibitem{STAR:sp}
\hrefCMSnoop {}{{STAR} Collaboration, ``Elliptic flow from two- and
  four-particle correlations in {Au+Au} collisions at {$\sqrtsNN = 130
  \GeV$}'',} \textit{ Phys. Rev. C} \textbf{ 66} (2002) 034904,
  \href{http://dx.doi.org/10.1103/PhysRevC.66.034904}{\doi{10.1103/PhysRevC.66.034904}},
\href{http://www.arXiv.org/abs/nucl-ex/0206001}{\texttt{arXiv:nucl-ex/0206001}}.

\bibitem{SP_ref}
\hrefCMSnoop {}{S.~A. Voloshin, A.~M. Poskanzer, and R.~Snellings, ``Collective
  phenomena in non-central nuclear collisions'',} \textit{
  Landolt-B{\"o}rnstein} \textbf{ 23} (2010) 293,
  \href{http://dx.doi.org/10.1007/978-3-642-01539-7_10}{\doi{10.1007/978-3-642-01539-7_10}},
  \href{http://www.arXiv.org/abs/0809.2949}{\texttt{arXiv:0809.2949}}.

\bibitem{Poskanzer:1998yz}
\hrefCMSnoop {}{A.~M. Poskanzer and S.~A. Voloshin, ``Methods for analyzing
  anisotropic flow in relativistic nuclear collisions'',} \textit{ Phys. Rev.
  C} \textbf{ 58} (1998) 1671,
  \href{http://dx.doi.org/10.1103/PhysRevC.58.1671}{\doi{10.1103/PhysRevC.58.1671}},
\href{http://www.arXiv.org/abs/nucl-ex/9805001}{\texttt{arXiv:nucl-ex/9805001}}.

\bibitem{Alt:2003ab}
\hrefCMSnoop {}{{NA49} Collaboration, ``Directed and elliptic flow of charged
  pions and protons in {Pb+Pb} collisions at {40A} and {158A GeV}'',} \textit{
  Phys. Rev. C} \textbf{ 68} (2003) 034903,
  \href{http://dx.doi.org/10.1103/PhysRevC.68.034903}{\doi{10.1103/PhysRevC.68.034903}},
\href{http://www.arXiv.org/abs/nucl-ex/0303001}{\texttt{arXiv:nucl-ex/0303001}}.

\bibitem{Qvec_ref}
\hrefCMSnoop {}{I.~Selyuzhenkov and S.~Voloshin, ``Effects of nonuniform
  acceptance in anisotropic flow measurements'',} \textit{ Phys. Rev. C}
  \textbf{ 77} (2008) 034904,
  \href{http://dx.doi.org/10.1103/PhysRevC.77.034904}{\doi{10.1103/PhysRevC.77.034904}},
  \href{http://www.arXiv.org/abs/0707.4672}{\texttt{arXiv:0707.4672}}.

\bibitem{np_D0}
\hrefCMSnoop {}{{CMS Collaboration}, ``Measurements of azimuthal anisotropy of
  nonprompt {D$^{0}$} mesons in {PbPb} collisions at 5.02 tev'',} \textit{
  Phys. Lett. B} \textbf{ 850} (2024) 138389,
  \href{http://dx.doi.org/10.1016/j.physletb.2023.138389}{\doi{10.1016/j.physletb.2023.138389}},
  \href{http://www.arXiv.org/abs/2212.01636}{\texttt{arXiv:2212.01636}}.

\bibitem{SP_ref_1}
\hrefCMSnoop {}{N.~Borghini and J.-Y. Ollitrault, ``Azimuthally sensitive
  correlations in nucleus-nucleus collisions'',} \textit{ Phys. Rev C} \textbf{
  70} (2004) 064905,
  \href{http://dx.doi.org/10.1103/PhysRevC.70.064905}{\doi{10.1103/PhysRevC.70.064905}},
  \href{http://www.arXiv.org/abs/nucl-th/0407041}{\texttt{arXiv:nucl-th/0407041}}.

\bibitem{lambdaC}
\hrefCMSnoop {}{{CMS Collaboration}, ``Study of charm hadronization with prompt
  {$\Lambda_{c}^{+}$} baryons in proton-proton and lead-lead collisions at
  {$\sqrtsNN=5.02\TeV$}'',} \textit{ JHEP} \textbf{ 01} (2024) 128,
  \href{http://dx.doi.org/10.1007/JHEP01(2024)128}{\doi{10.1007/JHEP01(2024)128}},
  \href{http://www.arXiv.org/abs/2307.11186}{\texttt{arXiv:2307.11186}}.

\bibitem{Ds-Meson-ALICE}
\hrefCMSnoop {}{{ALICE Collaboration}, ``Measurement of prompt {\PDpms} meson
  production and azimuthal anisotropy in {PbPb} collisions at
  {$\sqrtsNN=5.02\TeV$}'',} \textit{ Phys. Lett. B} \textbf{ 827} (2022)
  136986,
  \href{http://dx.doi.org/10.1016/j.physletb.2022.136986}{\doi{10.1016/j.physletb.2022.136986}},
  \href{http://www.arXiv.org/abs/2110.10006}{\texttt{arXiv:2110.10006}}.

\bibitem{He:2014tamu}
\hrefCMSnoop {}{M.~He, R.~J. Fries, and R.~Rapp, ``{Heavy flavor at the large
  hadron collider in a strong coupling approach}'',} \textit{ Phys. Lett. B}
  \textbf{ 735} (2014) 445,
  \href{http://dx.doi.org/10.1016/j.physletb.2014.05.050}{\doi{10.1016/j.physletb.2014.05.050}},
\href{http://www.arXiv.org/abs/1401.3817}{\texttt{arXiv:1401.3817}}.

\bibitem{hybrid-1}
J.~Casalderrey-Solana\hrefCMSnoop {}{ { et~al.}, ``A hybrid strong/weak
  coupling approach to jet quenching'',} \textit{ JHEP} \textbf{ 10} (2014)
  019,
  \href{http://dx.doi.org/10.1007/JHEP10(2014)019}{\doi{10.1007/JHEP10(2014)019}},
  \href{http://www.arXiv.org/abs/1405.3864}{\texttt{arXiv:1405.3864}}.

\bibitem{hybrid-2}
J.~Casalderrey-Solana\hrefCMSnoop {}{ { et~al.}, ``Angular structure of jet
  quenching within a hybrid strong/weak coupling model'',} \textit{ JHEP}
  \textbf{ 03} (2017) 135,
  \href{http://dx.doi.org/10.1007/JHEP03(2017)135}{\doi{10.1007/JHEP03(2017)135}},
  \href{http://www.arXiv.org/abs/1609.05842}{\texttt{arXiv:1609.05842}}.

\bibitem{hybrid-3}
J.~Casalderrey-Solana\hrefCMSnoop {}{ { et~al.}, ``Simultaneous description of
  hadron and jet suppression in heavy-ion collisions'',} \textit{ Phys. Rev. C}
  \textbf{ 99} (2019) 051901,
  \href{http://dx.doi.org/10.1103/PhysRevC.99.051901}{\doi{10.1103/PhysRevC.99.051901}},
  \href{http://www.arXiv.org/abs/1808.07386}{\texttt{arXiv:1808.07386}}.

\end{thebibliography}\endgroup

\cleardoublepage \appendix\section{The CMS Collaboration \label{app:collab}}\begin{sloppypar}\hyphenpenalty=5000\widowpenalty=500\clubpenalty=5000\cmsinstitute{Yerevan Physics Institute, Yerevan, Armenia}
{\tolerance=6000
A.~Belyaev\cmsorcid{0000-0003-1692-1173}, A.~Hayrapetyan, A.~Tumasyan\cmsAuthorMark{1}\cmsorcid{0009-0000-0684-6742}
\par}
\cmsinstitute{Institut f\"{u}r Hochenergiephysik, Vienna, Austria}
{\tolerance=6000
W.~Adam\cmsorcid{0000-0001-9099-4341}, J.W.~Andrejkovic, L.~Benato\cmsorcid{0000-0001-5135-7489}, T.~Bergauer\cmsorcid{0000-0002-5786-0293}, M.~Dragicevic\cmsorcid{0000-0003-1967-6783}, C.~Giordano\cmsorcid{0000-0001-6317-2481}, P.S.~Hussain\cmsorcid{0000-0002-4825-5278}, M.~Jeitler\cmsAuthorMark{2}\cmsorcid{0000-0002-5141-9560}, N.~Krammer\cmsorcid{0000-0002-0548-0985}, A.~Li\cmsorcid{0000-0002-4547-116X}, D.~Liko\cmsorcid{0000-0002-3380-473X}, M.~Matthewman, I.~Mikulec\cmsorcid{0000-0003-0385-2746}, J.~Schieck\cmsAuthorMark{2}\cmsorcid{0000-0002-1058-8093}, R.~Sch\"{o}fbeck\cmsAuthorMark{2}\cmsorcid{0000-0002-2332-8784}, D.~Schwarz\cmsorcid{0000-0002-3821-7331}, M.~Shooshtari\cmsorcid{0009-0004-8882-4887}, M.~Sonawane\cmsorcid{0000-0003-0510-7010}, W.~Waltenberger\cmsorcid{0000-0002-6215-7228}, C.-E.~Wulz\cmsAuthorMark{2}\cmsorcid{0000-0001-9226-5812}
% ...
\par}
\vskip\cmsinstskip
\dag:~Deceased\\
$^{1}$Also at Yerevan State University, Yerevan, Armenia\\
$^{2}$Also at TU Wien, Vienna, Austria\\
% ...
\end{sloppypar}
\end{document}